\documentclass[11pt]{article}
\usepackage{amsmath}
\usepackage{amssymb}
\usepackage[dvips]{graphicx}
\begin{document}
\title{On Multipolar Analytical Potentials for Galaxies}
\author{Daniel Vogt\thanks{e-mail: danielvt@ifi.unicamp.br}\\
Instituto de F\'{\i}sica Gleb Wataghin, Universidade Estadual de Campinas\\
13083-970 Campinas, S.\ P., Brazil
\and
Patricio S. Letelier\thanks{e-mail: letelier@ime.unicamp.br}\\
Departamento de Matem\'{a}tica Aplicada-IMECC, Universidade Estadual\\
de Campinas 13083-970 Campinas, S.\ P., Brazil}
\maketitle
\begin{abstract}
We present analytical potential--density pairs in three dimensions for the gravitational
field of galaxies, obtained by thickening the multipolar expansion
up to the quadrupole term. These may be interpreted as generalizations of the Miyamoto 
and Nagai potential--density pairs. With a suitable restriction on the possible values
of the multipole moments, the density distributions are positive and monotone
decreasing functions of the radial and axial coordinates.

\end{abstract}
\section{Introduction}

There are several three-dimensional analytical models in the literature for 
the gravitational field of different types of galaxies and galactic components.  
Jaffe (1983) and Hernquist (1990) discuss 
models for spherical galaxies and bulges. Three-dimensional models for flat galaxies
were obtained by Miyamoto and Nagai (1975) and Satoh (1980). de Zeeuw and Pfenniger (1989)
 considered a set of ellipsoidal models appropriate to 
galactic bars. Long and Murali (1992) derived simple potential--density pairs for a prolate and a triaxial bar by
softening a thin needle with a spherical and a Miyamoto and Nagai potential, respectively.
See Binney and Tremaine (1987) for a discussion of other galactic models. There also exist several general
relativistic models of disks, e.g., Morgan and Morgan (1969, 1970), Bi\v{c}\'{a}k et al.\ (1993), Lemos and Letelier (1994),
Gonz\'{a}lez and Letelier (2000, 2004), Vogt and Letelier (2003, 2005a). Recently, a general relativistic 
version of the Miyamoto and Nagai models was studied (Vogt and Letelier 2005b).
  
The potential--density pairs obtained by Miyamoto and Nagai (1975) are inflated versions of the thin-disk
family first derived by Toomre (1963). In this work we consider a set of three-dimensional 
potential--density pairs obtained by using the same procedure of Miyamoto and Nagai
on the multipolar expansion up to the quadrupole term. This generates a sequence of 
potential--density pairs that reduces to that of Miyamoto and Nagai for particular values of the
multipole moments. This is done in subsection \ref{subsec_1} and subsection \ref{subsec_2}.
The thin-disk limit is investigated in subsection \ref{subsec_3} and corresponds to generalizations
of Toomre's family of disks. Finally, the results are discussed in section \ref{sec_2}.
\section{Multipolar Models for Flattened Galaxies} \label{sec_1}

A general expression for a multipolar expansion in spherical coordinates can be cast as
\begin{equation}
\Phi=-\sum_{n=0}^{\infty} a_{n}\frac{P_{n}(\cos \theta)}{r^{n+1}} \mbox{,}
\end{equation} 
where $P_{n}$ are the Legendre polynomials and $a_{n}$ are coefficients related to the
multipolar moments. We consider only the expansion up to the
quadrupole term $(n=2)$. In cylindrical coordinates the explicit form reads
\begin{equation} \label{eq_1}
\Phi=-\frac{Gm}{\sqrt{R^2+z^2}}-\frac{Dz}{(R^2+z^2)^{3/2}}-
\frac{Q(-R^2+2z^2)}{2(R^2+z^2)^{5/2}} \mbox{,}
\end{equation}
where $D$ and $Q$ are the dipole and quadrupole moments, respectively. 

In the following three-dimensional models the mass-density distribution is obtained directly from Poisson equation,
\begin{equation} \label{eq_2}
\rho= \frac{1}{4\pi G} \left( \Phi_{,RR}+\frac{\Phi_{,R}}{R}+\Phi_{,zz} \right) \mbox{.}
\end{equation}
Other physical quantities of interest are the circular velocity, $v_{\mathrm{c}}$ of particles in the galactic
plane, the epicyclic frequency, $\kappa$, and the vertical frequency, $\nu$, of small oscillations
about the equilibrium circular orbit in the galactic plane. They are calculated with the expressions 
(Binney and Tremaine 1987):
\begin{align}
v_{\mathrm{c}}^2 &= R\Phi_{,R} \mbox{,} \label{eq_3} \\
\kappa^2 &= \Phi_{,RR}+\frac{3}{R}\Phi_{,R} \mbox{,} \label{eq_4} \\
\nu^2 &=\Phi_{,zz} \mbox{,} \label{eq_5}
\end{align}
where all quantities are evaluated on $z=0$. The stability conditions are set 
by $\kappa^2 \geq 0$ and $\nu^2 \geq 0$.

\subsection{Generalized Miyamoto and Nagai Model 2} \label{subsec_1}

We first consider $Q=0$, and apply the transformation $z \rightarrow a+\sqrt{z^2+b^2}$ on the
multipolar expansion equation (\ref{eq_1}), where $a$, $b$ are non-negative
constants. Using equation (\ref{eq_2}), we obtain
\begin{multline} \label{eq_6}
\tilde{\rho}=\frac{\tilde{b}^2}{4\pi\xi^3\left[\tilde{R}^2+(1+\xi)^2\right]^{7/2}} 
\left\{ \tilde{R}^4(1-\tilde{D}) 
+\tilde{R}^2(1+\xi)\left[ \tilde{D}(1-8\xi) \right. \right. \\
\left. \left. +(1+\xi)(2+3\xi)\right] +(1+\xi)^3\left[ 2\tilde{D}(1+4\xi)+(1+\xi)(1+3\xi) \right] \right\} \mbox{,}
\end{multline}
where the variables and parameters were rescaled in terms of $a$: $\tilde{R}=R/a$,
$\tilde{z}=z/a$, $\tilde{b}=b/a$, $\tilde{D}=D/(Gma)$, $\rho=m\tilde{\rho}/a^3$,
and $\xi=\sqrt{\tilde{z}^2+\tilde{b}^2}$.
For the particular value $\tilde{D}=1$, equation (\ref{eq_6}) reduces to the mass density of
the Miyamoto--Nagai model 2.

\begin{figure}
\centering
\includegraphics[scale=0.75]{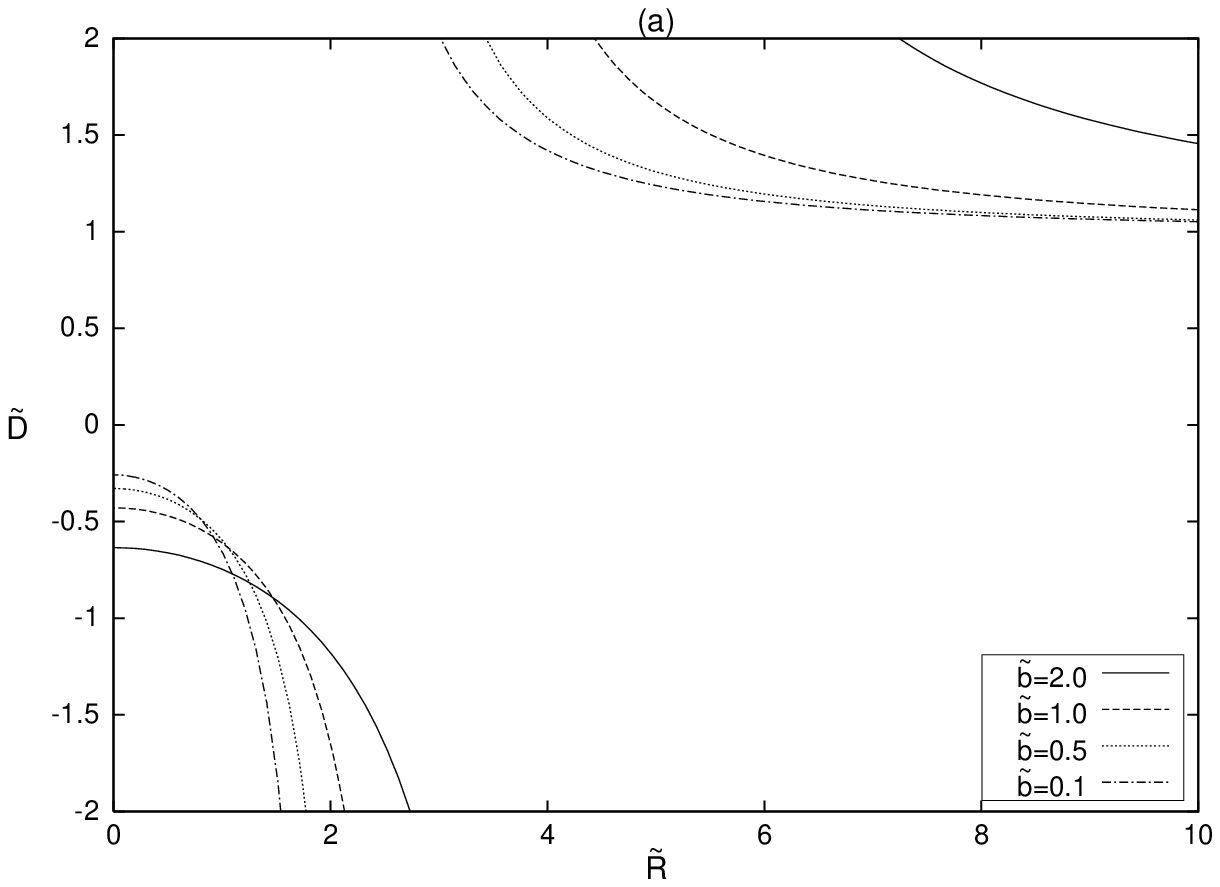}\\
\vspace{0.1cm}
\includegraphics[scale=0.75]{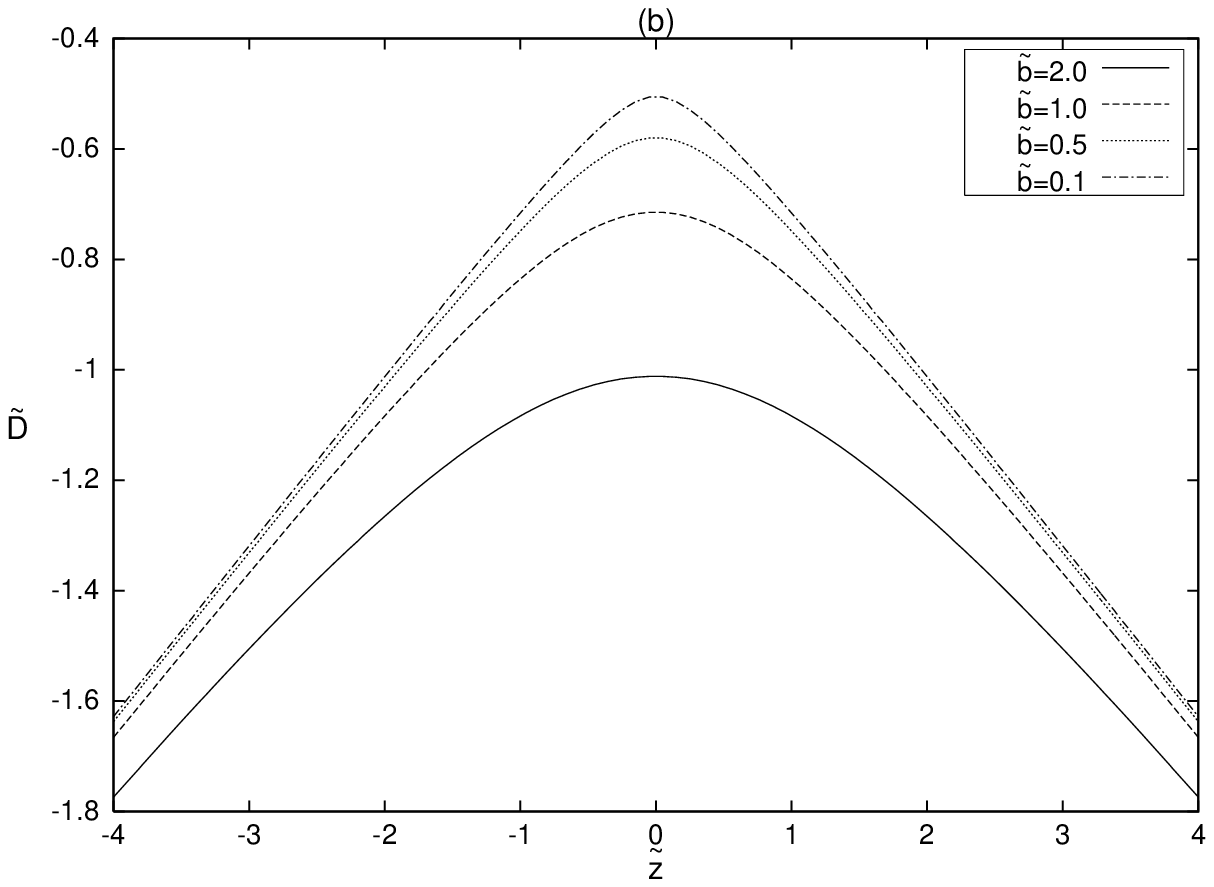}
\caption{(a) Curves of $\tilde{\rho}_{,\tilde{R}}=0$ as functions of $\tilde{R}$ and
$\tilde{D}$ for $\tilde{b}=2$, 1, 0.5, and 0.1. (b) $\tilde{\rho}_{,\tilde{z}}=0$
as functions of $\tilde{z}$ and $\tilde{D}$ for $\tilde{b}=2$, 1, 0.5, and 0.1.} \label{fig_1}
\end{figure}

Unfortunately the density distribution, equation (\ref{eq_6}), has some defects. For certain
ranges of the parameters $\tilde{b}$ and $\tilde{D}$ it is not a monotone decreasing
function of the radial and axial coordinates, and even has domains with negative densities.
To overcome this, we impose a reasonable restriction that the derivative of the density distribution
with respect to the $\tilde{R}$ coordinate along $\tilde{z}=0$ should have only one critical point at
$(\tilde{R},\tilde{z})=(0,0)$ as well as the derivative with respect to $\tilde{z}$ along $\tilde{R}=0$. 
A graphical study of the resulting equations is displayed in figures \ref{fig_1}a--b. In figure \ref{fig_1}a
the curves of $\tilde{\rho}_{,\tilde{R}}=0$ are plotted as functions of the radial
coordinate and the parameter $\tilde{D}$ for some values of $\tilde{b}$. In this case
decreasing $\tilde{b}$ narrows the interval of $\tilde{D}$ for which there are no other 
critical points, as on $\tilde{R}=0$. Figure \ref{fig_1}b shows the curve of
$\tilde{\rho}_{,\tilde{z}}=0$
as functions of $\tilde{z}$ and $\tilde{D}$ for the same values of $\tilde{b}$. One can see from both graphs
that the allowed interval for the parameter $\tilde{D}$ for a fixed value of $\tilde{b}$ in the case (a)
is contained in the allowed interval in case (b).  

The expressions for the circular velocity, epicyclic frequency and vertical frequency follow from 
equations (\ref{eq_3})--(\ref{eq_5})
\begin{align}
\tilde{v}_{\mathrm{c}}^2 &=\frac{\tilde{R}^2\left[ \tilde{R}^2+(1+\tilde{b})(1+\tilde{b}+3\tilde{D})
\right]}{\left[ \tilde{R}^2+(1+\tilde{b})^2 \right]^{5/2}} \mbox{,} \label{eq_7} \\
\tilde{\kappa}^2 &=\frac{1}
{\left[ \tilde{R}^2+(1+\tilde{b})^2 \right]^{7/2}}\left\{ \tilde{R}^4+\tilde{R}^2(1+\tilde{b}) 
\left[ 5(1+\tilde{b})-3\tilde{D} \right] \right. \notag \\
& \left. +4(1+\tilde{b})^3(1+\tilde{b}+3\tilde{D}) \right\} \mbox{,} \label{eq_8} \\
\tilde{\nu}^2 &=\frac{\tilde{R}^2(1+\tilde{b}-\tilde{D})+(1+\tilde{b})^2
(1+\tilde{b}+2\tilde{D})}{\tilde{b}\left[ \tilde{R}^2+(1+\tilde{b})^2 \right]^{5/2}} 
\mbox{,} \label{eq_9}
\end{align}
where $v_{\mathrm{c}}^2=Gm\tilde{v}_{\mathrm{c}}^2/a$, $\kappa^2=Gm\tilde{\kappa}^2/a^3$, 
and $\nu^2=Gm\tilde{\nu}^2/a^3$. The condition $\tilde{v}_{\mathrm{c}}^2 \geq 0$ and the stability
conditions $\kappa^2 \geq 0$ and $\nu^2 \geq 0$ also impose restrictions on the possible values
of $\tilde{D}$ and $\tilde{b}$. We find that $\tilde{D} \geq -(1+\tilde{b})/3$ ensures the 
positivity of the circular velocity and of the square of the epicyclic frequency, whereas the square of the vertical
frequency is always non-negative if $\tilde{D} \geq -(1+\tilde{b})/2$. 
\begin{figure}
\centering
\includegraphics[scale=0.65]{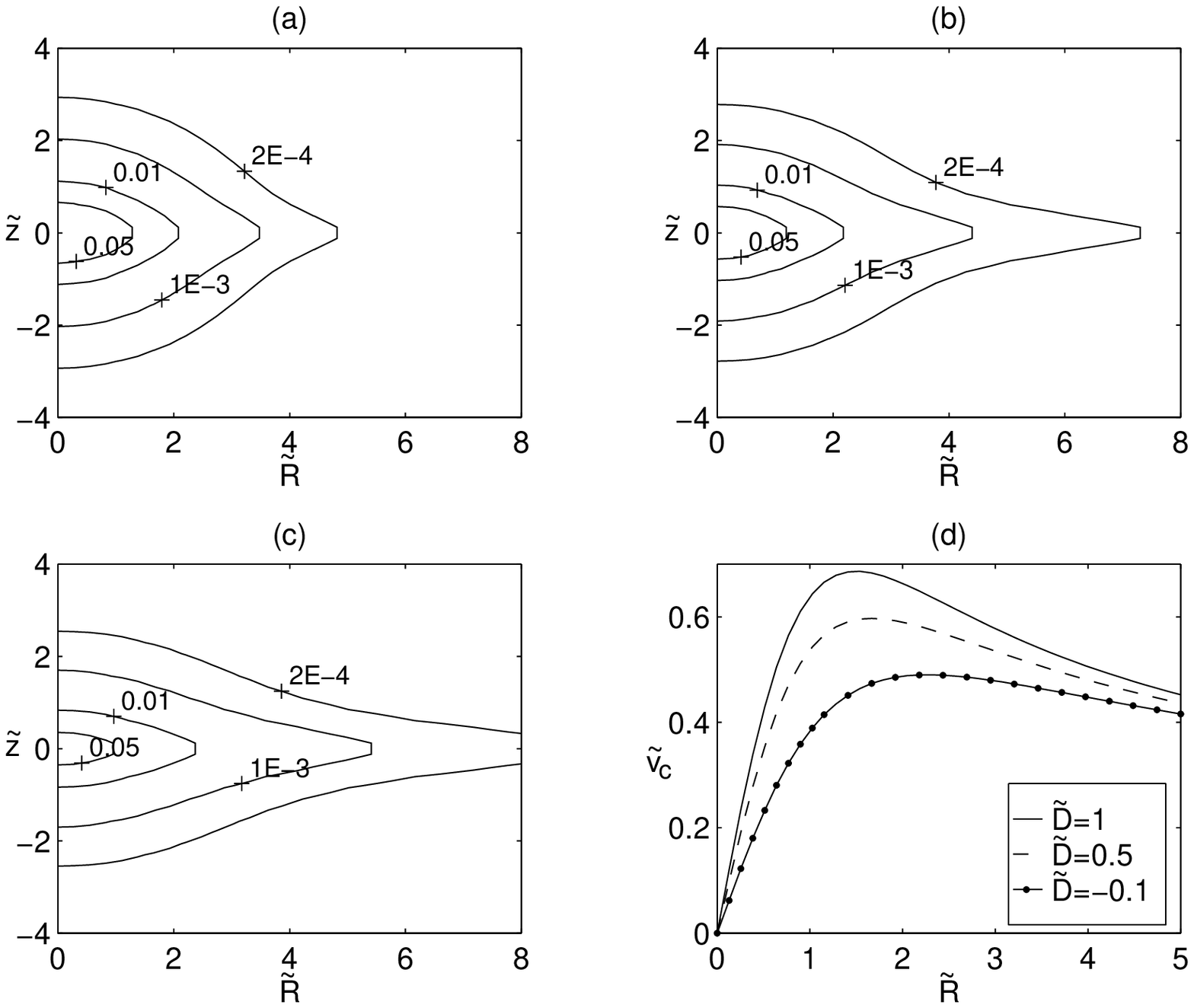}
\caption{Constant-density curves of equation (\ref{eq_6}) with parameters
$\tilde{b}=0.5$ and (a) $\tilde{D}=1$, (b) $\tilde{D}=0.5$, and (c) $\tilde{D}=-0.1$.
(d) The circular velocity $\tilde{v}_{\mathrm{c}}$ (equation (\ref{eq_7})), in the galactic plane for
cases (a)--(c).} \label{fig_2}
\end{figure}

\begin{figure}
\centering
\includegraphics[scale=0.65]{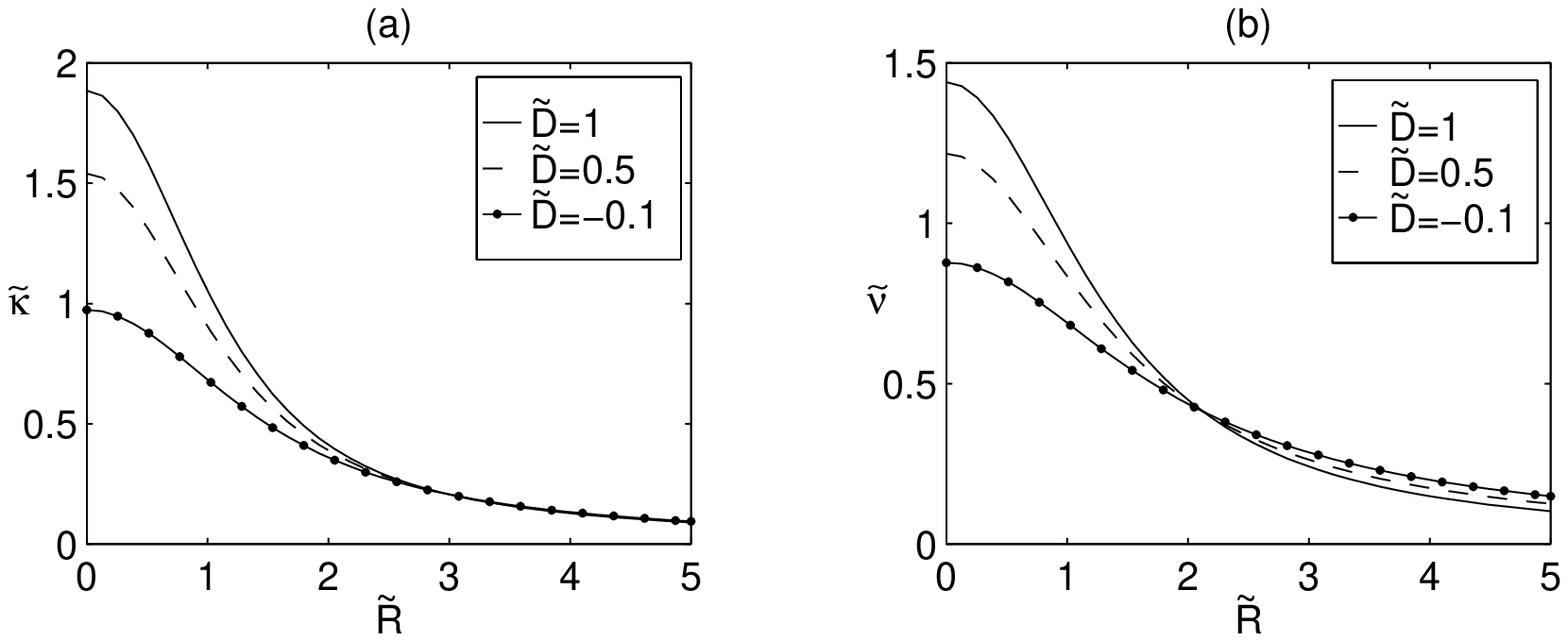}
\caption{Curves of (a) the epicyclic frequency, $\tilde{\kappa}$
(equation (\ref{eq_8})), and (b) the vertical frequency, $\tilde{\nu}$
(equation (\ref{eq_9})), for the same parameters as in figure \ref{fig_2}.} \label{fig_3}
\end{figure}
In figures \ref{fig_2}a--c we plot some isodensity curves of the density function,
equation (\ref{eq_6}), with parameter $\tilde{b}=0.5$ and (a) $\tilde{D}=1$, (b) $\tilde{D}=0.5$, and
(c) $\tilde{D}=-0.1$. For these values the density is a monotone decreasing function, as
can be checked from figure \ref{fig_1}. It is clearly seen that as the parameter $\tilde{D}$
decreases, the density distribution becomes more flattened. Figure \ref{fig_2}d and
figures \ref{fig_3}a--b
show, respectively, curves of the velocity profile, equation (\ref{eq_7}), the epicyclic frequency, equation (\ref{eq_8}),
and of the vertical frequency, equation (\ref{eq_9}), for the same parameters as in figures \ref{fig_2}a--c.
With decreasing $\tilde{D}$ the radius where the highest circular velocity occurs is increased and the 
epicyclic frequency is lowered. The vertical frequency becomes lower near the center, but for
$\tilde{R} \gtrapprox 2$ it is increased.
\subsection{Generalized Miyamoto and Nagai Model 3} \label{subsec_2}

We now consider the full expression equation (\ref{eq_1}) and apply the transformation
$z \rightarrow a+\sqrt{z^2+b^2}$. The resulting mass density distribution 
 reads
\begin{multline} \label{eq_10}
\tilde{\rho}= \frac{\tilde{b}^2}{8\pi\xi^3\left[\tilde{R}^2+(1+\xi)^2\right]^{9/2}} \left\{ 
2\tilde{R}^6(1-\tilde{D})
+3\tilde{R}^4 \left[ 2(1+\xi)^3-6\tilde{D}\xi(1+\xi) \right. \right. \\
\left. \left. -3\tilde{Q} \right] +3\tilde{R}^2(1+\xi)^2 \left[ 
2(1+\xi)^2(1+2\xi)+2\tilde{D}(1+\xi) - \tilde{Q}(1+25\xi) \right] \right. \\
\left. +2(1+\xi)^4 \left[ (1+\xi)^2(1+3\xi) +2\tilde{D}(1+\xi)(1+4\xi)+3\tilde{Q}(1+5\xi) \right] \right\}
\mbox{,}
\end{multline}
where the variables and parameters have been rescaled, as in subsection \ref{subsec_1} and $\tilde{Q}=
Q/(Gma^2)$. For particular values $\tilde{D}=1$, $\tilde{Q}=2/3$ we recover the Miyamoto and
Nagai model 3.

\begin{figure}
\centering
\includegraphics[scale=0.75]{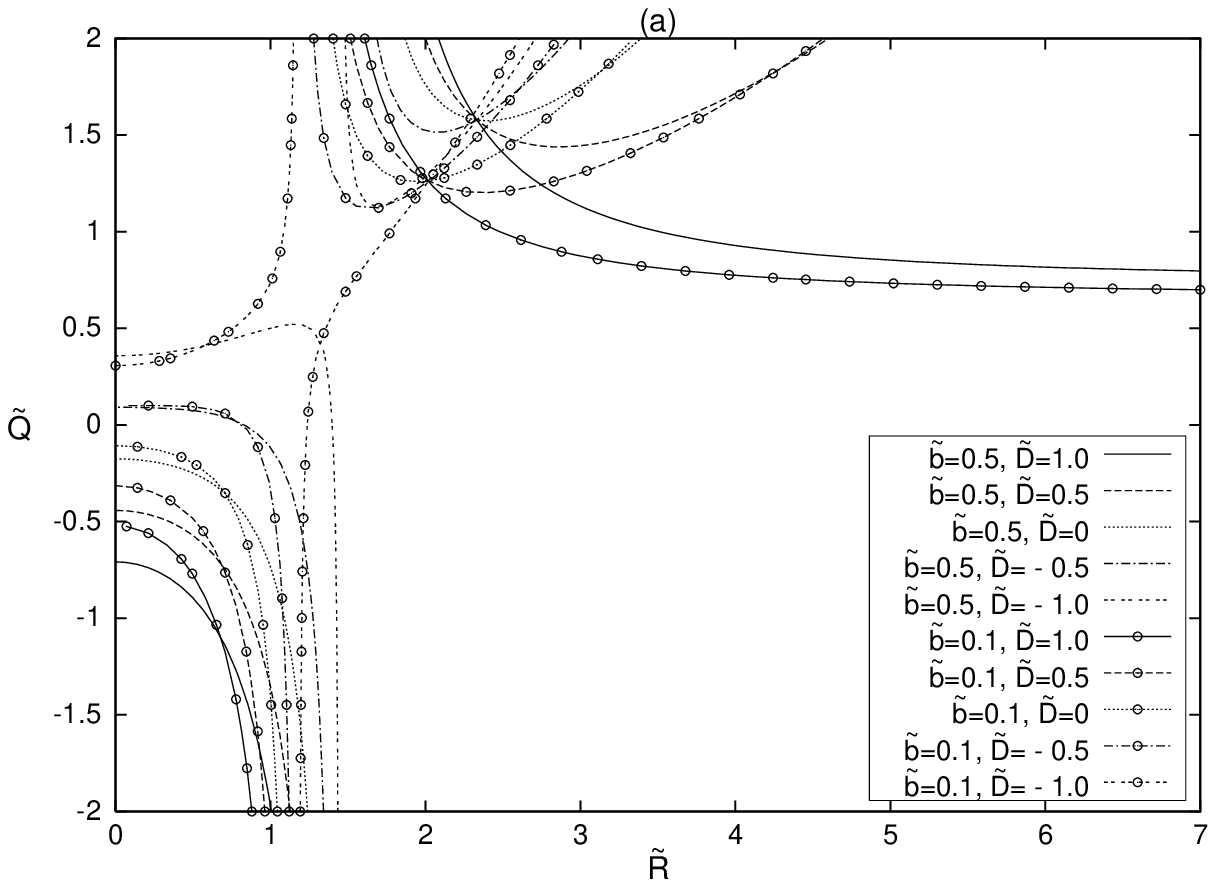}\\
\vspace{0.1cm}
\includegraphics[scale=0.75]{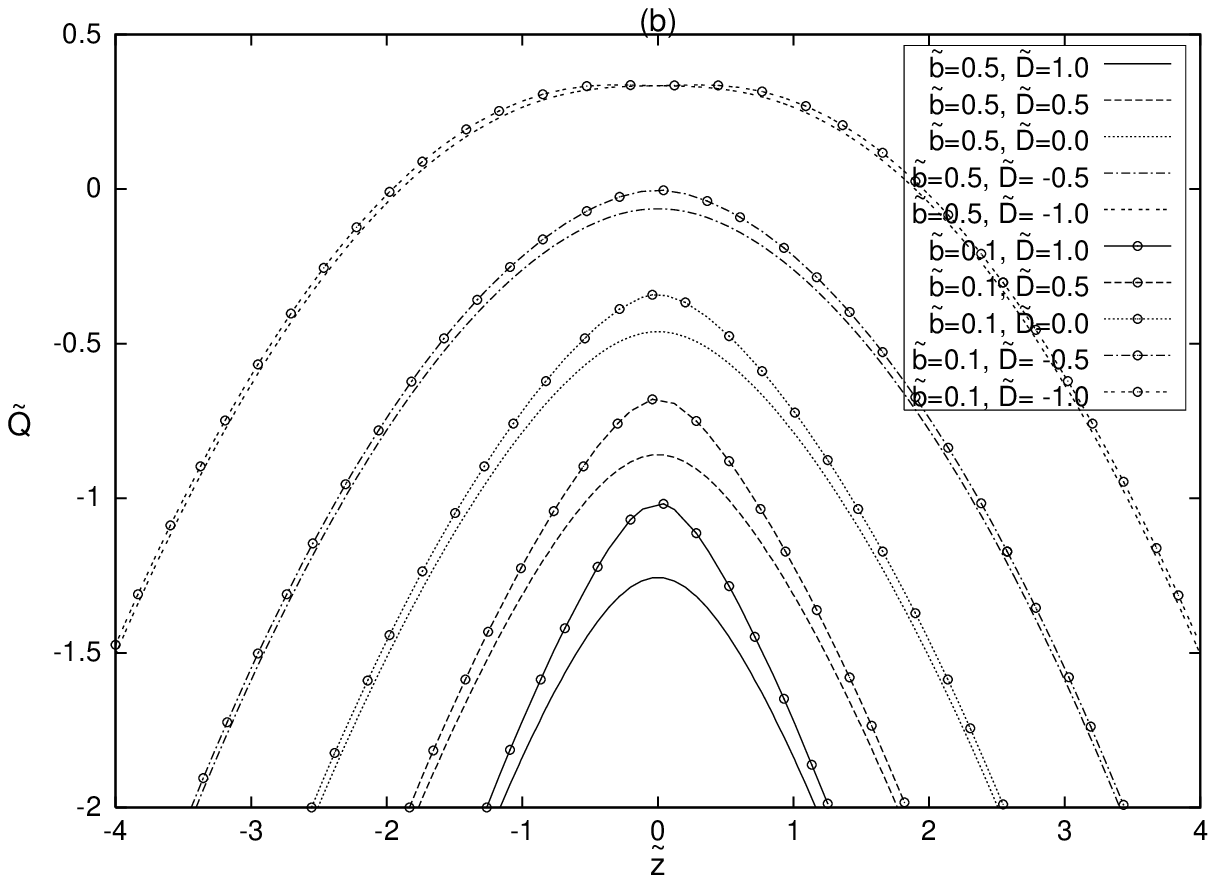}
\caption{(a) Curves of $\tilde{\rho}_{,\tilde{R}}=0$ as functions of $\tilde{R}$ and
$\tilde{Q}$ for some values of the parameters $\tilde{b}$ and $\tilde{D}$. (b) Curves of 
$\tilde{\rho}_{,\tilde{z}}=0$ as functions of $\tilde{z}$ and $\tilde{Q}$
for some values of the parameters $\tilde{b}$ and $\tilde{D}$.} \label{fig_4}
\end{figure}

Also here the range of the parameters must be restricted to produce physically acceptable 
density distributions. In figures \ref{fig_4}a--b we show, respectively,  some curves of $\tilde{\rho}_{,\tilde{R}}=0$
along $\tilde{z}=0$ and $\tilde{\rho}_{,\tilde{z}}=0$ along $\tilde{R}=0$ as functions of
$\tilde{Q}$ for some sets of values of $\tilde{D}$ and $\tilde{b}$. The expressions for the circular velocity, epicyclic frequency 
and vertical frequency follow from equations (\ref{eq_3})--(\ref{eq_5})
\begin{align}
\tilde{v}_{\mathrm{c}}^2 &=\frac{\tilde{R}^2}{2\left[ \tilde{R}^2+(1+\tilde{b})^2 \right]^{7/2}}
\left\{ 2 \tilde{R}^4 
+\tilde{R}^2 \left[ 4(1+\tilde{b})^2+6\tilde{D}(1+\tilde{b})-3\tilde{Q} \right] \right. \notag \\
& \left. +2(1+\tilde{b})^2 \left[(1+\tilde{b})^2+3\tilde{D}(1+\tilde{b})+6\tilde{Q} \right] \right\} \mbox{,}
\label{eq_11} \\
\tilde{\kappa}^2 &= \frac{1}{2\left[ \tilde{R}^2+(1+\tilde{b})^2 \right]^{9/2}}
\left\{ 2\tilde{R}^6 
+3\tilde{R}^4 \left[4(1+\tilde{b})^2-2\tilde{D}(1+\tilde{b})+\tilde{Q} \right] \right. \notag \\
& \left. +18\tilde{R}^2(1+\tilde{b})^2 \left[(1+\tilde{b})^2+\tilde{D}(1+\tilde{b})-3\tilde{Q} \right] \right. \notag \\
& \left. +8(1+\tilde{b})^4 \left[(1+\tilde{b})^2+3\tilde{D}(1+\tilde{b})+6\tilde{Q} \right] \right\} \mbox{,}
\label{eq_12} \\
\tilde{\nu}^2 &= \frac{1}{2\tilde{b}\left[ \tilde{R}^2+(1+\tilde{b})^2 \right]^{7/2}}
\left\{ 2\tilde{R}^4(1+\tilde{b}-\tilde{D}) +\tilde{R}^2(1+\tilde{b})
\left[4(1+\tilde{b})^2 \right. \right. \notag \\
& \left. \left. +2\tilde{D}(1+\tilde{b})-9\tilde{Q} \right] +2(1+\tilde{b})^3
\left[(1+\tilde{b})^2+2\tilde{D}(1+\tilde{b})+3\tilde{Q} \right] \right\} \mbox{,} \label{eq_13}
\end{align}
A graphical analysis of the constraints on the parameters imposed by $\tilde{v}_{\mathrm{c}}^2 \geq 0$, $\kappa^2 \geq 0$
and $\nu^2 \geq 0$ shows that they are contained in the restrictions imposed by the gradient of the mass density.
\begin{figure}
\centering
\includegraphics[scale=0.65]{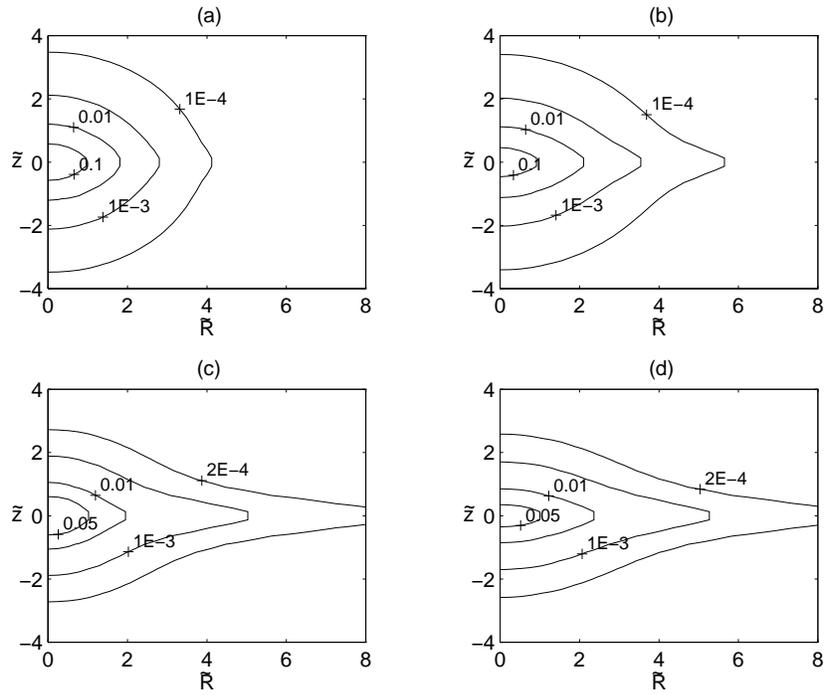}
\caption{Constant density curves of equation (\ref{eq_10}) with parameters
$\tilde{b}=0.5$ and (a) $\tilde{D}=1$, $\tilde{Q}=2/3$, (b) $\tilde{D}=1$, 
$\tilde{Q}=-0.1$, (c) $\tilde{D}=0$, $\tilde{Q}=2/3$, and
(d) $\tilde{D}=0$, $\tilde{Q}=-0.1$.} \label{fig_5}
\end{figure}

\begin{figure}
\centering
\includegraphics[scale=0.65]{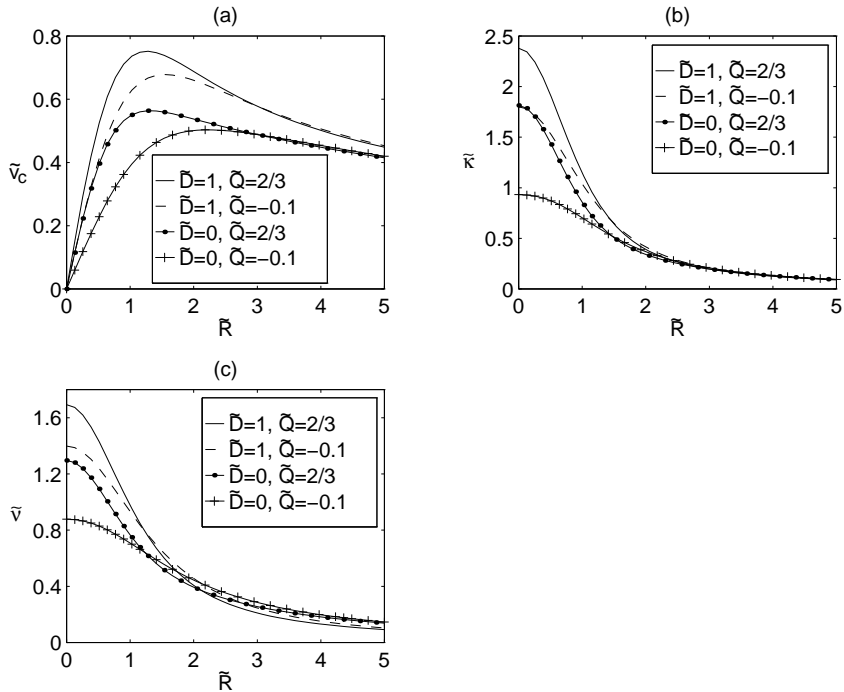}
\caption{Curves of (a) the circular velocity $\tilde{v}_{\mathrm{c}}$ equation (\ref{eq_11}) in the galactic plane,
(b) the epicyclic frequency $\tilde{\kappa}$ equation (\ref{eq_12}) and (c) the vertical frequency $\tilde{\nu}$ equation (\ref{eq_13}) for the same parameters as in figure \ref{fig_5}.} \label{fig_6}
\end{figure}

Some level curves of the mass density, equation (\ref{eq_10}), are displayed in figures \ref{fig_5}a--d.
In general, as $\tilde{D}$ and $\tilde{Q}$ are lowered for a fixed $\tilde{b}$, the mass distribution 
profile becomes flatter. We also note from figures \ref{fig_6}a--c that the maximum of the
circular velocity is shifted to larger radii, and the epicyclic frequencies and vertical frequencies
are lowered near the center. 
\subsection{Thin Disk Limit} \label{subsec_3}

\begin{figure} 
\centering
\includegraphics[scale=0.75]{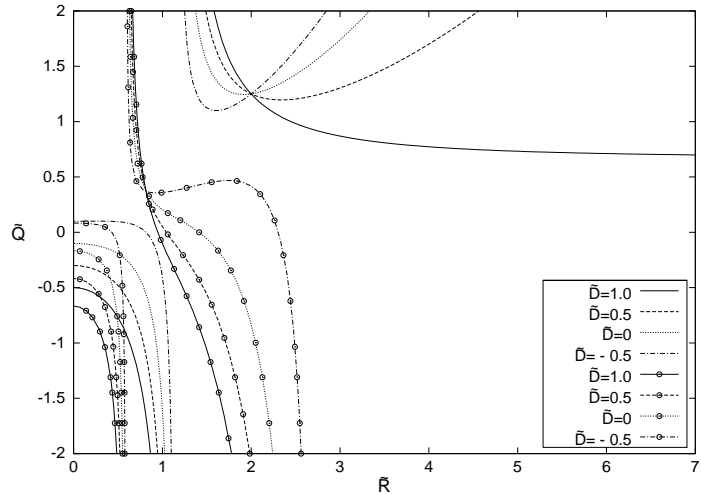}
\caption{Curves of $\tilde{\sigma}_{,\tilde{R}}=0$ (lines without circles) and of
$\tilde{\nu}=0$ (lines with circles) as functions of $\tilde{Q}$ and $\tilde{R}$ for $\tilde{D}=1$, 
0.5, 0, and -0.5.} \label{fig_7}
\end{figure}

To investigate the thin limit of the potential--density pairs deduced in the previous sections, it is more
convenient to rederive an expression for the surface density, $\sigma$, by applying the transformation
$z \rightarrow a+|z|$ on the multipolar expansion, equation (\ref{eq_1}) and using the well-known relation
(Binney and Tremaine 1987),
\begin{equation}
\sigma=\frac{1}{2\pi G} \Phi_{,z} \mbox{,}
\end{equation}
where the right-hand side is evaluated at $z \rightarrow 0^+$. We obtain
\begin{equation} \label{eq_14}
\tilde{\sigma}=\frac{1}{4\pi (\tilde{R}^2+1)^{7/2}} \left[ 2(\tilde{R}^2+1)^2(1-\tilde{D}) 
 +6\tilde{D}(\tilde{R}^2+1) -3\tilde{Q}(3\tilde{R}^2-2) \right] \mbox{.}
\end{equation}
The variables and parameters were rescaled as in subsection \ref{subsec_2}, and
$\sigma=m\tilde{\sigma}/a^2$. In particular cases with $\tilde{D}=1$, $\tilde{Q}=0$ and
$\tilde{D}=1$, $\tilde{Q}=2/3$ we obtain Toomre's models 2 and 3, respectively (Toomre 1963).
 Expressions for the circular velocity and epicyclic frequency 
are given by equation (\ref{eq_11}) and equation (\ref{eq_12}) with $\tilde{b}=0$, and the vertical
frequency is calculated from equation (\ref{eq_5}) evaluated at $ z \rightarrow 0^+$:
\begin{multline} \label{eq_15}
\tilde{\nu}^2=\frac{1}{2(\tilde{R}^2+1)^{9/2}} \left[ 2\tilde{R}^6+9\tilde{R}^4(2\tilde{D}-\tilde{Q}) \right. \\
\left. +6\tilde{R}^2(-1+\tilde{D}+12\tilde{Q}) -4(1+3\tilde{D}+6\tilde{Q}) \right] \mbox{.} 
\end{multline}
Figure \ref{fig_7} shows curves of $\tilde{\sigma}_{,\tilde{R}}=0$ (lines without circles) and
$\tilde{\nu}=0$ (lines with circles) as functions of $\tilde{Q}$ and $\tilde{R}$ for some values of the parameter $\tilde{D}$.
In the thin limit, the disks always have regions of vertical instability. The Kuzmin disk $(\tilde{D}=\tilde{Q}=0)$ is
stable for $\tilde{R} \geq \sqrt{2}$; Toomre's model 2 $(\tilde{D}=1$, $\tilde{Q}=0)$ is stable 
for $\tilde{R} \geq \sqrt{2}\sqrt{-2+\sqrt{6}}$; if $\tilde{Q}=0$ the disk is
stable for
\begin{equation}
\tilde{R}  \geq \frac{\sqrt{2}}{2} \left[ 1-9\tilde{D}+\sqrt{3(27\tilde{D}^2+2\tilde{D}+3)} \right]^{1/2}
\end{equation}
if $\tilde{D} \geq -1/3$; and
\begin{align}
0 & \leq \tilde{R} \leq \frac{\sqrt{2}}{2} \left[ 1-9\tilde{D}-\sqrt{3(27\tilde{D}^2+2\tilde{D}+3)} \right]^{1/2}, \\
& \text{ and } \tilde{R}  \geq \frac{\sqrt{2}}{2} \left[ 1-9\tilde{D}+\sqrt{3(27\tilde{D}^2+2\tilde{D}+3)} \right]^{1/2}
\end{align}
if $\tilde{D} < -1/3$.
\begin{figure}
\centering
\includegraphics[scale=0.65]{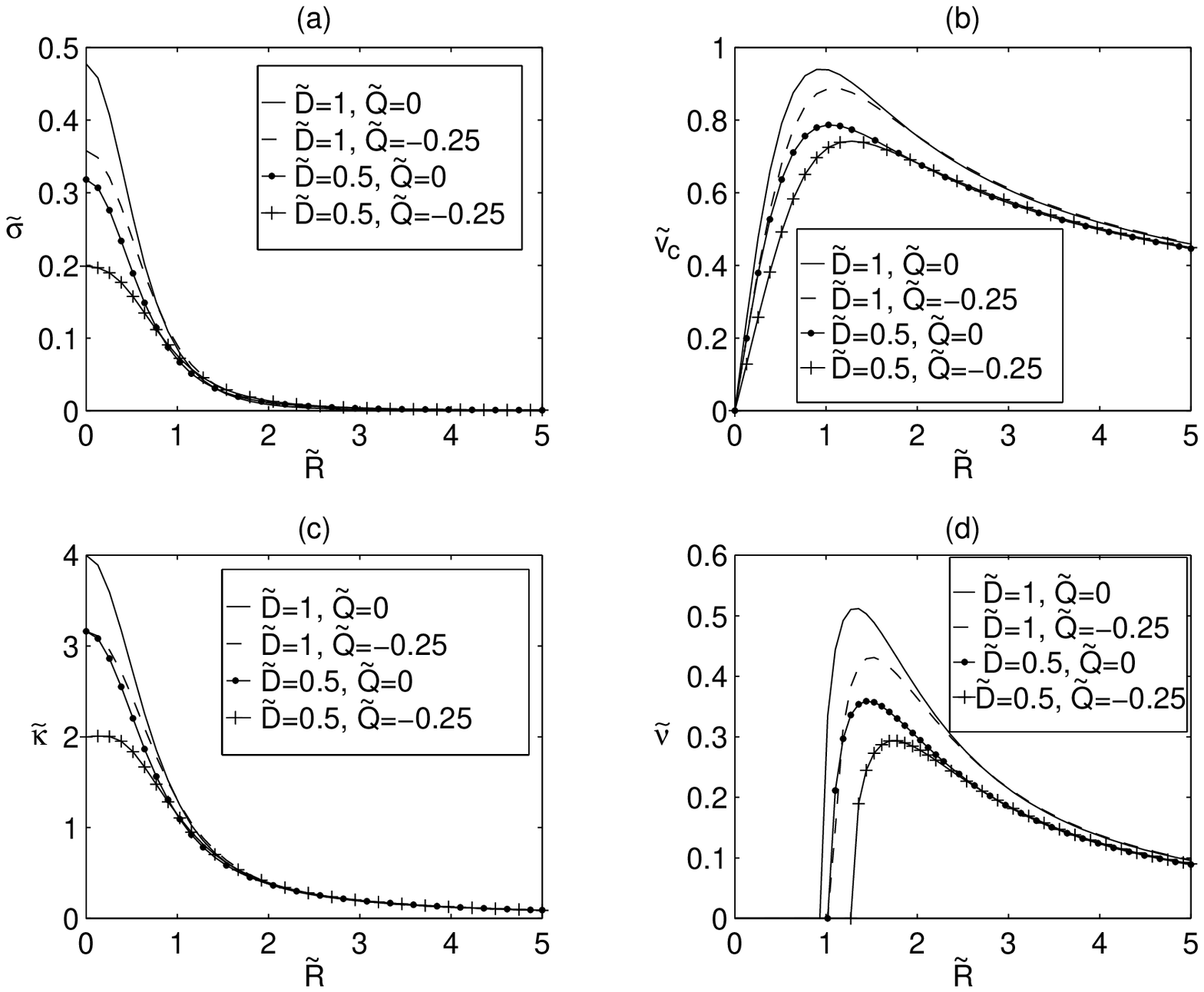}
\caption{(a) Curves of the surface density equation (\ref{eq_14}), (b) the circular velocity equation (\ref{eq_11}) with $\tilde{b}=0$, (c)  the epicyclic frequency equation (\ref{eq_12}) with $\tilde{b}=0$
and the vertical frequency equation (\ref{eq_15}). Parameters: $\tilde{D}=1$, $\tilde{Q}=0$ (solid lines);
$\tilde{D}=1$, $\tilde{Q}=-0.25$ (dashed lines); $\tilde{D}=0.5$, $\tilde{Q}=0$ (lines with points); 
$\tilde{D}=0.5$, $\tilde{Q}=-0.25$ (lines with +).} \label{fig_8}
\end{figure}

In figure \ref{fig_8} we plot some curves of the surface density equation (\ref{eq_14}), the
circular frequency equation (\ref{eq_11}) with $\tilde{b}=0$, the epicyclic frequency
equation (\ref{eq_12})
also with $\tilde{b}=0$ and the vertical frequency equation (\ref{eq_15}) for some values of
$\tilde{D}$ and $\tilde{Q}$. The behaviour of the three first physical quantities, displayed as the multipole parameters
are decreased, is very similar to those of the thick-disk model of the previous section.

\section{Discussion} \label{sec_2}

A Miyamoto and Nagai-like transformation was applied on the multipolar
expansion up to the quadrupole term to produce potential--density pairs for
flattened galaxies. These models, the first without the quadrupole term and the second with the
quadrupole term, may be viewed as generalizations of the Miyamoto and
Nagai models 2 and 3, respectively. The thin disk limit was also investigated
and corresponds to generalizations of Toomre's models 2 and 3.  For each model we also
calculated the velocity profile in the galactic plane and the epicyclic and vertical
frequencies of oscillation of perturbed circular orbits.

The major drawback of our models is that the density distribution is not a monotone 
decreasing function of the radial and axial coordinates for arbitrary values of the free 
parameters. Thus we imposed the condition that the derivatives of the density distributions 
with respect to $R$ and $z$ do not have critical points, except at the origin. These imposed
restrictions on the possible ranges of the multipole moments. We found that, except in the
thin case limit, those restrictions were also suficient to ensure non-negative circular velocities
and also non-negative epicyclic and vertical frequencies (disk stability on the $z=0$ plane).

\bigskip
D.\ Vogt thanks CAPES for financial support. P.\ S.\ Letelier thanks CNPq and FAPESP for financial support. This research has made use of NASA's Astrophysics Data System.
\section*{References}

Bi\v{c}\'{a}k, J., Lynden-Bell, D., \& Katz, J. 1993, Phys. Rev. D,  47, 4334 \\
Binney, S., \& Tremaine, S. 1987, Galactic Dynamics (Princeton, Princeton University Press), p. 121 \\
de Zeeuw, T., \& Pfenniger, D. 1988, MNRAS,  235, 949 \\
Gonz\'{a}lez, G. A., \& Letelier, P. S. 2000, Phys. Rev. D, 62, 064025 \\
Gonz\'{a}lez, G. A., \& Letelier, P. S. 2004, Phys. Rev. D, 69, 044013 \\
Hernquist, L. 1990, ApJ,  356, 359 \\
Jaffe, W. 1983, MNRAS,  202, 995 \\
Lemos, J. P. S., \& Letelier, P. S. 1994, Phys. Rev. D, 49, 5135 \\
Long, K., \& Murali, C. 1992, ApJ,  397, 44 \\
Miyamoto, M., \& Nagai, R. 1975, PASJ  27, 533 \\
Morgan, L., \& Morgan, T. 1970, Phys. Rev. D, 2, 2756 \\
Morgan, T., \& Morgan, L. 1969, Phys. Rev., 183, 1097 \\
Satoh, C. 1980, PASJ,  32, 41 \\
Toomre, A. 1963, ApJ, 138, 385 \\
Vogt, D., \& Letelier, P. S. 2003, Phys. Rev. D, 68, 084010 \\
Vogt, D., \& Letelier, P. S. 2005a, Phys. Rev. D, 71, 084030 \\
Vogt, D., \& Letelier, P. S. 2005b, MNRAS, 363, 268 \\
\end{document}